\documentclass[11pt]{article}
\usepackage[dvips]{graphicx}
\begin{document}
%\tighten
%\draft

\title{\Large \bf Supernova-Neutrino Studies with $^{100}$Mo}
 
\author{\\
H.~Ejiri$^1$, J.~Engel$^2$, and N.~Kudomi$^3$\\
{\small $^1$IIAS, Kizu-cho, Kyoto, 619-0225; 
JASRI-Spring8, Mikazuki-cho, Hyogo, 679-5198}\\
{\small $^2$Department of Physics and Astronomy, CB3255, University of 
North Carolina,}\\
{\small Chapel Hill NC 27599}\\
{\small $^3$RCNP, Osaka-University, Ibaraki, Osaka 567-0047}
}

\date{}

\maketitle

\begin{abstract}

{\normalsize We show that supernova neutrinos can be studied by 
observing their charged-current interactions with $^{100}$Mo, which has 
strong spin-isospin giant resonances. Information about both the 
effective temperature of the electron-neutrino sphere and the 
oscillation into electron neutrinos of other flavors can be extracted 
from the electron (inverse $\beta$) spectrum.   We use measured hadronic 
charge-exchange spectra and the Quasiparticle Random Phase Approximation 
to calculate the charged-current response of $^{100}$Mo to electron 
neutrinos from supernovae, with and without the assumption of 
oscillations.  A scaled up version of the MOON detector for $\beta 
\beta$ and solar-neutrino studies could potentially be useful for 
spectroscopic studies of supernova neutrinos as well.}\\

\vspace{.1cm}


\end{abstract} 

\vspace{.1cm}

Neutrinos carry away most of the energy from core-collapse supernovae.
Supernova neutrinos (SN-$\nu$'s) can be observed on the earth, and their
spectrum contains information about conditions inside the supernova as well
as their own properties.  Here we aim to show that $^{100}$Mo, which
responds strongly to spin-isospin probes, is useful for studying supernova
weak processes and SN-$\nu$ oscillations, and that a good SN-$\nu$ detector
can be realized by scaling up the proposed $\beta\beta$ and solar-neutrino
detector MOON.

Though there is much we don't know about supernovae, the consensus of
modelers is that SN-$\nu$'s are released roughly thermally from the
supernova remnant after diffusing to the surface of last scattering, called
the ``neutrino sphere".  They therefore escape with an energy corresponding
approximately to the thermal energy spectrum at the sphere
\cite{bet90,woo94,mil93}.  In this picture there are really three neutrino
spheres, one for electron neutrinos ($\nu_e$'s), one for electron
antineutrinos ($\bar{\nu}_e$'s), and one for the other flavors ($\nu_x$'s
and $\bar{\nu}_x$'s).  The $\nu_e$ sphere has the largest radius of these
because $\nu_e$'s interact with matter via both charged- and neutral-current
reactions.  So do $\bar{\nu}_e$'s, but the excess of neutrons over protons
in the supernova remnant means that they scatter less frequently through
charged-current interactions, so that the radius of their neutrino sphere is
smaller.  The other neutrinos ($\nu _x$, $\bar {\nu }_x$ ), with only the
neutral-current interactions, decouple deeper within the star.  Since the
temperature in the supernova core increases as the radius gets smaller,
these last neutrinos will have the highest energy, and the $\nu _e$'s the
lowest energy.

The SN-$\nu$ spectrum for a given neutrino species is thought to be 
roughly 
\cite{woo94} \cite{mil93}
\begin{equation} 
S(E_{\nu}) =
cT^{-1}_{\nu}\frac{(E_{\nu}/T_{\nu})^2}{{\rm exp}(E_{\nu}/T_{\nu}-a) + 
1}
\end{equation} 
where $T_{\nu}$ is the temperature at the neutrino sphere, $a$ is the
degeneracy parameter, and $c$ is a normalization constant.  Numerical
simulations can be approximately reproduced with temperatures $T_{\nu}$
of about 3.5 MeV for $\nu_e$'s, 5 MeV for $\bar{\nu}_e$'s, and 8 MeV for
$\nu _x$'s and $\bar{\nu} _x$'s, with the degeneracy parameter $a$ taken
to vanish.  Accordingly, the average $\nu$ energies are $<E(\nu _e)>
\sim$11 MeV, $<E(\bar{\nu} _e)> \sim$16 MeV, $<E(\nu _x)> \sim$25 MeV,
and the spread of SN-$\nu$ energies covers the wide region of $E \sim$
5-70 MeV.  

Measuring the $\nu_e$ spectra would provide us information on the electron
neutrino sphere, and thus tell us if our supernova models are on the right
track.  It could also tell us about neutrino oscillations; if our ideas
about where the neutrinos leave the supernova are correct, $\nu_e$'s with
energies above 30 MeV or so are rarely emitted directly from the supernova.
An excess of high-energy $\nu_e$'s reaching the earth would be strong
evidence for oscillations from $\nu_x$ to $\nu_e$.  

A number of detectors can study neutrinos in the event of a nearby
supernova.  They have the ability to detect either the charged-current
$\nu_e$ ($\bar{\nu}_e$) interaction, which produces electrons (positrons),
or the neutral current interaction (for all flavors), which usually results
in the production of neutrons and photons, or both.  Antineutrinos from
SN1987A were observed by the Kamiokande \cite{hir87} and IMB \cite{bio87}
groups in water Cerenkov detectors via the reaction $p+ \bar{\nu}_e
\rightarrow n + e^+$.  SuperKamkiokande, with multi tons of water, and the
Sudbury Neutrino Observatory (SNO), with kilotons of heavy water, are
powerful detectors for SN-$\nu$'s (see ref.\ \cite{bea98}).
SuperKamiokande, however, has a high threshold ($Q \sim 15$ MeV) for the
charged-current interaction of $\nu_e$'s with $^{16}$O.  The effective
threshold energy, including a 5-MeV threshold for detecting an electron
produced by the charged-current interaction, is therefore about 20 MeV, well
above the average energy of neutrinos emitted from the $\nu_e$ sphere.  As a
result, while the detector is good for charged-current $\bar{\nu}_e$ 
interactions, it will have a hard
time saying anything about the flux or energy distribution of thermally
emitted $\nu_e$'s.  Detectors based on liquid scintillator, such as KamLAND
\cite{kam}, also have a high threshold for $\nu_e$ charged current
interactions with $^{12}$C --- about 17 MeV, with an effective threshold
energy of around 20 MeV.  They will not be able to study neutrinos from the
$\nu_e$ sphere either.  SNO, on the other hand has a low threshold, plus
the eventual ability to separately measure charged and neutral current
interactions.

Ref.\ \cite{ful99} shows that information on SN-$\nu$ energies and
oscillations can be obtained by measuring the number of neutrons produced by
neutrino scattering from heavy nuclei.  The method is very good for getting
gross features of the SN-$\nu$ spectra and possible oscillations, and the
proposed facilities OMNIS \cite{cli97}, SBNO \cite{smi97}, and LAND
\cite{har96} are based largely on the detection of neutrons.  These
detectors cannot easily measure the spectra of charged-current events,
however.  In addition, the $\nu_e$ cross section on lead is small at low
energies because of the extreme concentration of Gamow-Teller(GT) strength
in a single resonance at high excitation, so that information about
low-energy neutrinos will be hard to obtain.

A low-threshold charged-current detector would therefore add to our ability
to study neutrinos from the $\nu_e$ sphere, particularly if the detector 
could measure the spectrum of
electrons from the neutrino interactions and if it were made of a material
with a large SN-$\nu$ cross section.  If our ideas about the $\nu_e$ sphere
are grossly wrong, such a detector would also tell us that.  By looking for
high-energy $\nu_e$'s, the detector could also complement existing and
planned facilities in studying SN-$\nu$ oscillations.

A recent paper \cite{eji00} argues that MOON (Mo Observatory Of Neutrinos),
containing a few tons of $^{100}$Mo, would be useful for studies of both
$\beta\beta$ decay (having the ability to detect a neutrino mass as low as
$<m_{\nu }>\sim $0.03 eV) and real time studies of low energy solar-$\nu$
spectra.  In what follows we discuss how $^{100}$Mo and a scaled-up MOON
would be useful for studying SN-$\nu $'s as well as low energy solar-$\nu
$'s.

The isotope $^{100}$Mo has a threshold ($Q$ value) for the 
charge-exchange process 
\begin{equation}
\label{eq:Mo}
\nu_e + ^{100}{\rm Mo} \longrightarrow e^- + ^{100}{\rm Tc}
\end{equation}
of only $Q$=0.17 MeV, much less than other detectors with light nuclei 
such as $^{12}$C and $^{16}$O. In addition, one expects $^{100}$Mo to 
exhibit a large response to charged-current interaction of SN-$\nu$'s 
because of the  large neutron excess (isospin $T_z \equiv (N-Z)/2 = 8$), 
which enhances the strengths of spin-isospin giant resonances. 
 
Recent measurements of $^{100}$Mo($^3$He,t)$^{100}$Tc cross sections
\cite{aki97} confirm this expectation.  They show that at energies below 50
MeV this reaction (changing neutrons to protons) primarily excites four
isospin giant resonances \cite{eji00a}:  the isobaric analog resonance (IAR)
with $J^{\pi}=0^+$, the Gamow-Teller giant resonance (GTR) with
$J^{\pi}=1^+$, the isovector dipole resonance (IDR) with $J^{\pi}=1^-$, and
the isovector spin-dipole resonance (ISDR) with $J^{\pi}=0^-, 1^-, 2^-$.
The GTR is accompanied by a low-energy shoulder (GTR') below the main peak.
The IAR and IDR are excited by operators in coordinate space (times the
isospin-raising operator $\tau _+$) while the GTR and ISDR involve the spin
operator $\vec{\sigma}$ as well.

The strength in these resonances are spread over the excitation energy
region 5-35 MeV, with the centroid of IAR at 11.6 MeV, the GTR and GTR'
centroids at 13.4 MeV and 8 MeV, and the centroid of the combined dipole
resonances, which cannot be separated by the experiment, at 21 MeV
\cite{aki97}.  This energy range corresponds nicely with that of SN-$\nu$'s,
which will therefore also proceed primarily through the resonances,
particularly the GTR.  The spread of the GT strength down to below 5 MeV
together with the low $Q$ value of the charge-exchange process in eq.\
(\ref{eq:Mo}) make the effective threshold as low as a few MeV, well below
the average SN-$\nu_e$ energy.  As we discuss next, we can actually use the
measured charge-exchange response to calibrate a calculation of SN-$\nu$
cross sections.

Precise expressions for the matrix elements that govern these cross sections
are given in Ref.\ \cite{neu}.  We use the charge-changing quasiparticle
random phase approximation (QRPA) to calculate most of these matrix
elements.  Our approach is similar to that of ref.\ \cite{eng} with
improvements such as a larger model space (about 20 single-particle levels
around the Fermi surface for both protons and neutrons), and a better
treatment of the Coulomb interaction of the outgoing electron \cite{engcou}
The interaction we use has the same $\delta$-function form, with parameters
adjusted to fit the observed GTR energy and the low-lying spectrum in
$^{100}$Mo.  For neutrinos of the energies we consider here, it is
sufficient to include multipoles up to $J=4$.

In the important $1^+$ channel, we replace the QRPA calculation with the
measured GT strength.  Because the neutrino cross section in this channel is
determined mainly by the operator $j_0(qr) \vec{\sigma} \tau_+$, rather than
the GT operator $\vec{\sigma} \tau_+$, we must supplement the measured GT
strength with a q-dependent form factor.  We obtain the form factor from the
Helm model \cite{helm} , which takes the strength to be peaked at the
nuclear surface.  We cannot repeat this procedure for higher multipoles
because they are not separated in the measured spin-isospin dipole strength
distributions (and the overall normalization is not known).  Our theoretical
strength distributions, however, reproduce the measured ones quite well, up
to the unknown normalization constant.  We choose not to artificially quench
the strength of the dipole transitions because no clear evidence supports
such quenching; muon capture, in fact, argues against it \cite{mucap}.  The 
use experimental data to calibrate these calculation should make them 
accurate to within a factor of two at worst\footnote{Our cross sections for 
the highest-energy neutrinos may be slightly too small because of the 
restrctions on our single-particle space.}.

Fig.\ 1 shows the calculated cross section for $\nu_e$ scattering on
$^{100}$Mo as a function of neutrino energy.  The charge-changing
flux-averaged SN-$\nu$ cross sections, broken down by multipole, appear in
Table 1.  We consider two cases, non-oscillating SN-$\nu_e$'s, and
SN-$\nu_{x}$'s (either $\nu_{\mu}$'s or $\nu_{\tau}$'s, but not both) that
oscillate completely into $\nu_e$'s.  We label these two cases by $\nu_e$
and $\nu_{ex}$.  GT-like transitions, the major part of the $1^+$
contribution, dominate the cross section, particularly for the
non-oscillating $\nu_e$'s, which have lower energy on average.  The cross
section for $\nu_{ex}$'s is more than an order of magnitude larger than that
for non-oscillating $\nu_e$'s.  Since $\nu_x$'s have energies well above the
GT and dipole giant resonances, the phase space for $\nu_{ex}$ scattering is
quite large.

Fig.\ 2 shows the calculated spectra (or counts per MeV ton of 
$^{100}$Mo) of electrons produced by the charged-current interactions of 
both $\nu_e$ and $\nu _{ex}$ from a typical supernova 10 kpc away, 
emitting $3 \times 10^{53}$ ergs.  We assume that the SN energy is 
partitioned equally among all neutrino flavors.  The average electron 
energy of 25 MeV for $\nu _{ex}$ is about 2.5 times larger than the 
average energy of 11 MeV for $\nu_e$, reflecting the ratio of temperatures 
at the two neutrino spheres.  This means that the flux of $\nu_e$'s is 
higher by the same 
factor, a fact reflected in the count rates.  

The large electron energy for $\nu _{ex}$, together with the large cross
section, make a $\nu _{ex}$ component clearly visible; the observation
of a large fraction of the events at relatively high electron energies
would be a clear signal of oscillations.  But the figure also tells us
about the importance of a low threshold.  In a large enough detector,
the neutrinos from the $\nu_e$ sphere will clearly be observable if
there are no oscillations.  If there is a resonant effect that converts
all $\nu_e$'s into $\nu_x$'s then, of course, no detector will tell us
anything about the $\nu_e$ sphere.  But if --- as in the
solution to the solar and atmospheric neutrino problems with large 
$\theta_{12}$ and $\theta_{13}=0$ --- half of
the emitted $\nu_e$'s oscillate into $\nu_x$'s, the number of events
from $\nu_e$ relative to that from $\nu_{ex}$ will be the same as shown
in the figure.  At energies below 10 or 15 MeV, a significant fraction
of the events would come therefore from the $\nu_e$ sphere, and one
could learn something about the spectrum of emitted $\nu_e$'s even in the
presence of oscillations.

How large a detector would we need?  Our calculations imply that with a
supernova 10 kpc away emitting $3 \times 10^{53}$ ergs, one would detect
about 2 $\nu_e$'s and about 13 $\nu _{ex}$'s (under the no-oscillation and
maximum-oscillation scenarios discussed above) in a detector with 30 tons of
$^{100}$Mo.  Such a detector is roughly equivalent to the MOON detector
discussed in ref.\ \cite{eji00}, which contains 3.3 tons of $^{100}$Mo,
corresponding to 34 tons of natural molybdenum; we argue below that the 
cross sections on other
molybdenum isotopes will be of the same order as in $^{100}$Mo.  Thus, even
as proposed MOON could conclusively answer the question of whether there are
oscillations from $\nu_x$ to $\nu _e$.  One would need a detector at least
an order of magnitude larger, however, to look closely at the spectrum of
non-oscillating $\nu_e$'s, and thus the characteristics of the electron
neutrino sphere.

The proposed MOON detector could be realized either as a supermodule of
plastic scintillators with thin natural or enriched molybdenum layers or
a liquid scintillator doped with natural or enriched molybdenum.  The
former design can be scaled up to a kiloton of natural molybdenum by
increasing the Mo thickness of the modules from of 0.03 g/cm$^2$ to 2
g/cm$^2$ ($\sim$1mm).  The average energy loss in the foil is only 1.7
MeV for the electron from ($\nu_e,e$).  Thus the effective threshold
energy ($Q$ value + detector threshold energy) could still as low as 2
MeV, far below the average energy of the $\nu_e$'s.

The cross-section of SN-$\nu_e$'s per unit weight for $^{100}$Mo is about as
large as that for $^{208}$Pb because of the large neutron excess $(N-Z)/A$ =
0.16 and the small thereshold energy.  What are the effects of using natural
molybdenum rather than $^{100}$Mo?  As Table 1 suggests, the non-oscillating
$\nu_e$'s mainly excite the GT resonance, so that their cross sections are
very roughly given by the product of the GT strength $B(GT)$ and a phase
space factor $G$.  The GT strength is roughly proportional to $T_z$ and $G$
is proportional to $(E_{\nu}-Q_{G})^2$, where $E_{\nu}$ is the effective
neutrino energy and $Q_{G}$ the $Q$ value for exciting the GT resonance.
$Q_G$ has a slight linear dependence on $T_z$ \cite {eji00a} \cite {hor81}.
These facts imply that the use of natural Mo with the $T_z \sim $6 (on
average) will reduce the $\nu_e$ count rate by something on the order of
35$\%$ from the rate in $^{100}$Mo, which has $T_z=8$.  The $\nu_{ex}$'s
excite all the resonances discussed above, but the strength associated with
those also depends linearly on $T_z$.  If we assume that the energies of
those resonances scale the same way as that of the GT resonance, we find
that the count rates in natural molybdenum for the high-enegy neutrinos are
perhaps 30$\%$ smaller than in $^{100}$Mo.  The $Q$ values for the ground
state transitions are just a few MeV higher for other Mo isotopes than for
$^{100}$Mo.  Thus a detector with natural Mo can still have a low effective
threshold, and efficiencies of the same order as those with $^{100}$Mo.
Such a detector could therefore serve our purpose:  providing useful
information about the spectrum at the electron-neutrino sphere, as well as
observing oscillations and measuring the effective temperature at the
$\nu_x$ sphere.  And if our ideas about the emission of neutrinos by
supenovae are wrong, the detector would be sensitive enough to tell us so.

Mo, which has a large neutron excess, is not so sensitive to antineutrinos
because most of the GT transitions are Pauli blocked.  Neutral-current
interactions of SN-$\nu$'s would excite the Mo isotopes, which decay mostly
by emitting neutrons and successive $\gamma$ rays.  These particles deposit
energy in a large volume of scintillator, and so could be separated from
charged-current events, which have a single electron signal
accompanied by several neutron and $\gamma $ signals.  But other detectors,
such as SK and SNO, would see more neutral current events (and many more
antineutrino-charged-current events) than this one would \cite{bea98,lan96}.
A Mo detector, with its sensitivity to $\nu_e$'s, would therefore not 
obviate other detectors, but would complement them nicely.  Information on 
the antineutrino spectrum, for example, could strengthen evidence for 
oscillations that might be observed in the neutrino spectrum.

In summary, $^{100}$Mo and other Mo isotopes have large cross sections
for SN-$\nu_e$ and SN-$\nu_{ex}$.  A scaled up version of
MOON, which could measure electron energy spectra down to $\sim 2$
MeV, would be useful both for studying neutrino oscillations and for
learning about conditions at the electron-neutrino sphere.  With the
exception of SNO, which has an effective threshold of few MeV, no 
other detector could do the latter as well.  Other
heavy nuclei with large $N - Z$ could conceivably be used in place of
molybdenum in the liquid-scintillator version of the detector.\\

We thank Professors R.G.H.\ Robertson and P.\ Vogel for valuable 
discussions. We were supported in part by the U.S. Department of Energy 
under grant DE--FG02--97ER41019.

\newpage
\begin{table}[t]
\begin{center}
\caption{Calculated flux-averaged neutrino cross sections in units of 
$10^{-41}$ cm$^2$, with contributions from each multipole given 
separately}
\label{t:1}
\vspace{0.5cm}
\begin{tabular}{|lcc|}
\hline\hline
& $\nu_e$ & $\nu_{ex} $ \\ \hline
$0^+$ &0.65 &8.94  \\
$0^-$ &0.02 &0.59  \\
$1^+$ &4.62 &32.34  \\
$1^-$ &0.14 &11.86  \\
$2^+$ &0.04 &4.62  \\
$2^-$ &0.34 &14.00  \\
$3^+$ &0.03 &3.78  \\ 
$3^-$ & --- &1.00 \\
$4^+$ & --- &0.23 \\
$4^-$ & --- &0.79 \\\hline
total&5.84 &78.16 \\ \hline\hline
\end{tabular}
\end{center}
\end{table}

\newpage

\begin{figure}[t!]
\begin{center}
\includegraphics[angle=0,width=10cm]{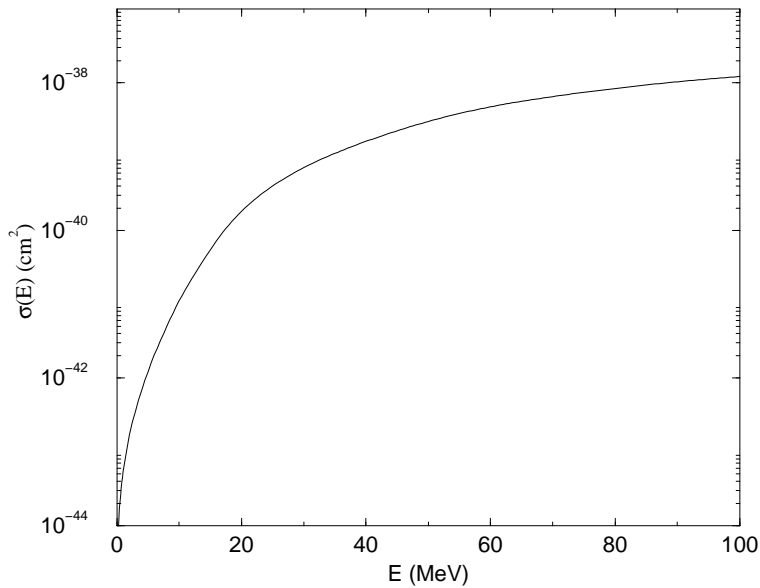}
\end{center}
\caption[Figure 1]{The calculated cross section for $\nu_e$ charged-current 
scattering on $^{100}$Mo, as a function 
of neutrino energy.}

\label{f:1}
\end{figure}

\begin{figure}[hb]
\begin{center}
\includegraphics[angle=0,width=10cm]{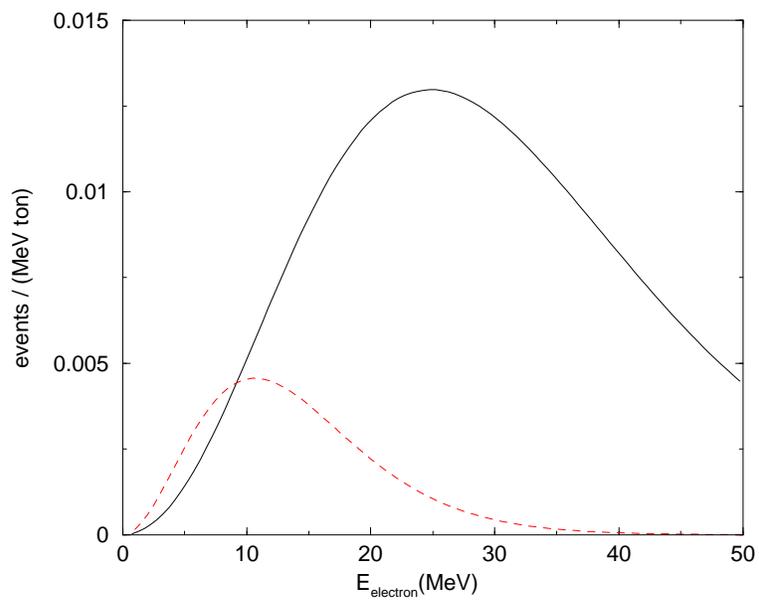}
\end{center}
\caption[Figure 2]{The calculated energy spectra of electrons produced 
by charged-current interactions of both $\nu_e$ (dashed line) and 
$\nu_{ex}$ (solid line), assuming equipartition of SN
energy among all flavors.  The vertical axis is the number of electrons 
per MeV per ton of $^{100}$Mo.}

\label{f:2}
\end{figure}

\end{document}